\documentstyle[epsf,graphicx]{mn}
\newcommand{\mincir}{\raise
-2.truept\hbox{\rlap{\hbox{$\sim$}}\raise5.truept\hbox{$<$}\ }}
\newcommand{\magcir}{\raise
-2.truept\hbox{\rlap{\hbox{$\sim$}}\raise5.truept\hbox{$>$}\ }}
\newcommand{\minmag}{\raisei
-2.truept\hbox{\rlap{\hbox{$<$}}\raise6.truept\hbox{$<$}\ }}

\title[Shapes of Poor Groups]{The Shape of Poor Groups of Galaxies}
\author[M.Plionis, S.Basilakos, H.M. Tovmassian]
{M. Plionis$^{1,2}$, S. Basilakos$^{2}$, H.M. Tovmassian$^{1}$ \\
\vspace{0.2cm}
$^{1}$ Instituto Nacional de Astrof\'isica \'Optica y 
Electr\'onica, AP 51 y 216, 72000, Puebla, Pue, M\'exico \\
$^{2}$ National Observatory of Athens, I. Metaxa \& B. Pavlou, Lofos Koufou,
Palaia Penteli, 15236, Athens, Greece}

\begin{document}

\maketitle

\begin{abstract}
We estimate the distribution of intrinsic  shapes of  UZC-SSRS2
groups of galaxies from the distribution of their apparent shapes. 
We measure the projected group axial ratio using 
the moments of their discrete galaxy distribution.
Then using the non-parametric kernel method to estimate the smooth
apparent axial ratio distribution we numerically invert a set of integral
equations to recover the corresponding intrinsic distribution 
under the assumption that groups are either
oblate or prolate spheroids.
We find that the prolate spheroidal model fits very well the  UZC-SSRS2
group distribution with a true mean axial ratio 
$\langle \beta \rangle \simeq 0.3$ and $\sigma_{\beta}\simeq 0.15$. 
This shows that groups of
galaxies are significantly more elongated, both on the plane of the
sky and in 3 dimensions, than clusters of galaxies.
The poorest groups that we consider, those with 4 members, 
are even more elongated than the overall population with 85\% of the
groups having $\beta \mincir 0.4$. 
\end{abstract}

\begin{keywords}
galaxies: galaxies: groups: general 
\end{keywords}

\section{INTRODUCTION} 
In Cold Dark Matter models, structure formation evolves in a hierarchical
fashion with aggregation of smaller mass units along large-scale
anisotropic structures. 
Since virialization processes tend to sphericalize initial anisotropic
distributions of matter, the shape of cosmic structures is
related to their formation processes and evolutionary stage
and thus it is extremely important to unambiguously determine their 
intrinsic shapes.

Apart from disk galaxies all cosmic structures on larger scales
appear to be dominated by prolate like shapes. This has been shown to
be the case for clusters of galaxies (cf. 
Carter \& Metcalfe 1980; Plionis, Barrow \& Frenk 1991; 
Cooray 1999; Basilakos, Plionis \& Maddox 2000) as well as 
for superclusters which show a predominance
of filamentary like shapes both observationally, theoretically
 and in Cosmological
N-body simulations (cf Zeldovich, Einasto \& Shandarin 1982;
Shandarin \& Zeldovich 1983;
Broadhurst et al. 1990; de Lapparent, Geller \& Huchra 1991;
Plionis, Jing \& Valdarnini 1992;
Jaaniste et al 1998; Sathyaprakash et al. 1998; 
Valdarnini, Ghizzardi \& Bonometto 1999; Basilakos, Plionis \&
Rowan-Robinson 2001).
In the case of Hickson or Shakhbazian compact groups it has been shown that
they are even flatter than clusters and most probably prolate-like
configurations with typical true axial ratios $\sim 0.3$
(Vardanian \& Melik-Alaverdian 1978; Hickson et al. 1984; 
Malykh \& Orlov 1986; Oleak et al. 1995).

It is obvious that the intrinsic shape of cosmic
structures can be lost when projected  on the plane of the sky and
therefore it is important to deal with such and  
other systematic effects that can hide the true
shape of cosmic structures. 
Different studies have attempted to recover the distribution of
intrinsic shapes from the corresponding apparent distribution
using inversion techniques based on the assumption that their
orientations are random.

The plan of the paper is the following:
In Section 2 we describe the group sample that we use. We attempt to
identify the extent of projection contamination and we describe
the projected shape determination method.
In Section 3 we invert the projected axial ratio
distribution and recover the corresponding intrinsic one.
The discussion and our conclusions are presented in Section 4.

\section{Projected Poor Group Shapes}
We use the recent UZC-SSRS2 group catalogue (Ramella et al 2002) which
is based on the Updated Zwicky Catalogue (UZC; Falco et al. 1999) and
the Southern Sky Redshift Survey (SSRS2; da Costa et al 1998), to 
measure the projected group shape distribution and hence attempt to 
estimate their intrinsic shape. The catalogue has a sky coverage of 
4.69 sr, it is limited to $m_{B} \simeq 15.5$ and contains 1168 groups
of galaxies exceeding a number density contrast threshold of $\delta
\rho/\rho=80$. The group catalogue was constructed using the
well-known friends of friends algorithm with a linking
parameter that scales with increasing redshift, 
in order to take into account the galaxy selection function. 

In order to have a representative sample of the true underlying
group population, we have chosen to study those groups the number
density of which, within some limiting redshift, is relatively constant. 
We have derived the group number density as a function of redshift in
equal volume shells (see Fig. 1) 
and we have found that it is roughly
constant out to $cz\sim 5500$ km s$^{-1}$, which we choose as our
limit. Within this velocity limit we have a total of 245 groups
with 4 or more members. 
\begin{figure}
\mbox{\epsfxsize=8.7cm \epsffile{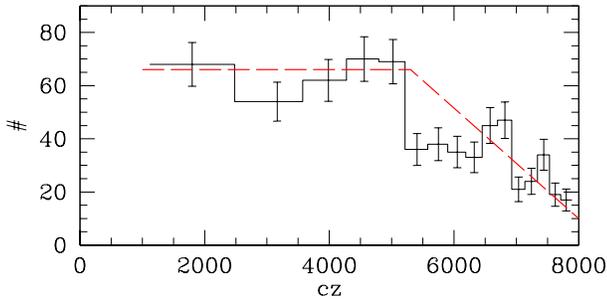}}
\caption{The group number density in equal volume shells. Note that we
plot the number density only for $cz> 1800$ km s$^{-1}$. 
At smaller distances there is a local excess in detected groups.
The dashed line shows a crude fit to the data 
}
\end{figure}

\begin{figure}
\mbox{\epsfxsize=8.7cm \epsffile{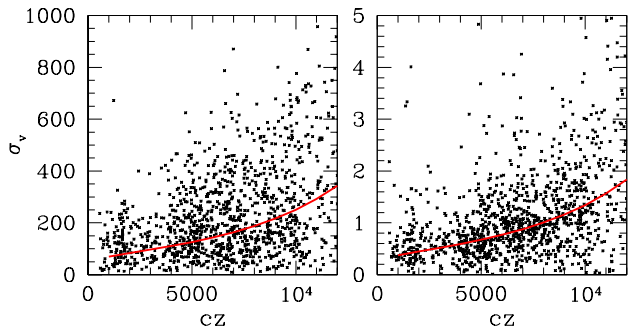}}
\caption{The dependence of the group velocity dispersion (left panel)
and maximum intergalaxy separation in $h^{-1}$ Mpc (right panel) 
on the group redshift. The continuous lines represent the dependence
of the line-of-sight velocity ($V_L$; left panel) 
and the projected separation ($D_L$; right panel)  
galaxy pair linking parameters on the redshift.}
\end{figure}

In Fig. 2 we present the group's velocity dispersion as well as their maximum
intergalaxy separation as a function of redshift. A strong $z$-dependence is
evident, although such dependence is rather weak within $\sim 5500$ km
s$^{-1}$. What is the cause of this redshift trend ?

We note that in order for the Ramella et al. (2002) algorithm to
identify groups having the same limiting density contrast at the
different distances, it was necessary to take into account the
drop with redshift of the galaxy space density which is due to the
magnitude limit of the parent galaxy catalogue. The way this is
accomplished is by increasing with redshift 
the linking parameters, used to identify the group members.
The Ramella et al (2002) algorithm, like most
others, uses two linking parameters; one that links pairs of 
galaxies below some projected separation, ie., $D_{12}<D_L$, and one that links
galaxies with line-of-sight velocity differences below so threshold,
ie., $V_{12} (\equiv |V_1-V_2|) <V_L$. 
Therefore in order to keep the limiting density enhancement of
the detected groups constant it is necessary to scale the linking 
parameters by a
distant dependent quantity that compensates, as discussed before, for
the drop of the selection function of the parent galaxy
population. For example, Ramella et al (2002)
 use the following scaling:
$$D_L=D_o R \;\;\;\;\; {\rm and} \;\;\;\;\; V_L=V_o R\;,$$ 
where
$$R=\left(\int_{L_{min}}^{L_{max}} \Phi(L) {\rm d}L / 
\int_{L_{min}(r)}^{L_{max}} \Phi(L) {\rm d}L \right)^{1/3}$$
with $\Phi(L)$ the galaxy luminosity function, $L_{min}$ the faintest
luminosity at which galaxies with the limiting magnitude of the
catalogue can be visible at the fiducial velocity used (1000 km/sec)
and $L_{min}(r)$ the faintest luminosity at which galaxies with the 
limiting magnitude of the catalogue can be visible at the distance
$r$.
In Figure 2 we also plot by continuous lines the values of $V_L$ (multiplied
by a factor 1/5) and $D_L$ (multiplied by 1.5) 
in the corresponding plots, which evidently follow the redshift trend 
observed in the data. 

The apparently good correlation between the
increase with redshift of the size and velocity dispersion of groups 
and the corresponding increase of the group linking parameters
suggests that the bias is introduced by the algorithm itself,
especially by the way it deals with the drop of the galaxy redshift 
selection function. 
The fact that groups of galaxies are clustered (eg. 
Zandivarez, Merch\'an \& Padilla 2003 and references therein)
implies that the increase with redshift of the group linking parameters may 
increasingly include, as part of the identified groups, galaxies 
belonging to neighbouring structures; a fact which will affect 
both the size and velocity dispersion of the resulting groups. 
This systematic bias could be quantified with the use of simulations 
but it is out of the scope of the present work.

We conclude that the probability that the groups identified constitute 
a homogeneous and biased free sample as well as that they constitute 
real dynamical entities, decreases with redshift.

\subsection{Shape determination methods}
We have used two methods to determine the group shapes.
The first one is according to Rood (1979), by which the axial ratio, 
$q\equiv(b/a)$,
is such that $a$ is the angular distance between the most widely
separated galaxies in the group, and $b$ is the the sum of the angular 
distances $b_{1}$ and $b_{2}$ of the most distant galaxies 
on either side of the line $a$ joining the most separated
galaxies. The second one is the moments of inertia method 
(cf. Basilakos et al 2000 and references therein).
In Fig. 3 we compare the results of the two methods. It is evident that
they both provide equivalent axial ratios, except in a few
cases where the position of the 
largest pair separation, used to determine $a$ in the first method, 
is strongly asymmetrical with respect to the rest of the galaxies of
the group (the outliers in the figure).
\begin{figure}
\mbox{\epsfxsize=8.7cm \epsffile{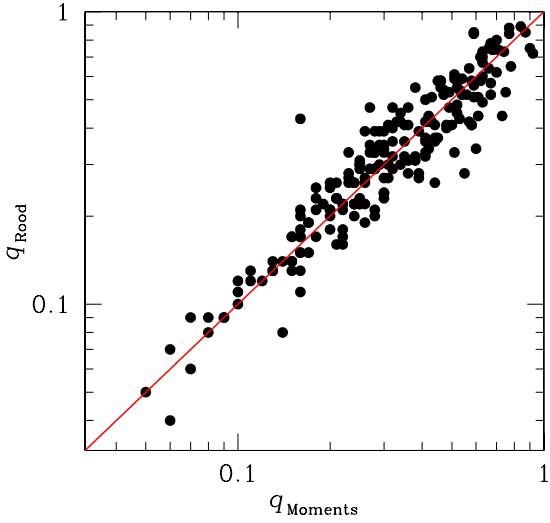}}
\caption{Comparison of the group axial ratio determined using the
moments of inertia and Rood's (1979) methods.}
\end{figure}

In what follows we will use the results of the more robust moments of
inertia method.
Due the the slightly more complicated nature of this
method, we describe it in more details below. 
Firstly, the galaxy equatorial positions are transformed into an equal area
coordinate system, centered on the group center of mass, using:
$x =(\alpha_{g}-\alpha_{gr}) \times \cos(\delta_{gr})$ and
$y =\delta_{g}-\delta_{gr}$, where subscripts $g$ and $gr$ refer to
galaxies and the groups, respectively. 
We then evaluate the moments:
\begin{eqnarray}
I_{11} & = & \sum_{i} w_{i}(r_{i}^{2}-x_{i}^{2}) \nonumber \\
I_{22} & = & \sum_{i} w_{i}(r_{i}^{2}-y_{i}^{2}) \nonumber \\
I_{12} & = & I_{21}=-\sum_{i} w_{i}x_{i}y_{i}
\end{eqnarray}
with $w_{i}$ the statistical weight of each point (in our case
$w_i=1$) and $r_i$ the distance of the $i^{th}$ galaxy from the group
center of mass. 
Note that because the inertia tensor is symmetric we have
$I_{12}=I_{21}$. Diagonalizing the inertia tensor
\begin{equation}\label{eq:diag}
{\rm det}(I_{ij}-\lambda^{2}M_{2})=0 \;\;\;\;\; {\rm (M_{2} \;is \; 
2 \times 2 \; unit \; matrix.) }
\end{equation}
we obtain the eigenvalues $\lambda_{1}$, $\lambda_{2}$, from which we
define the principal axial ratio of the configuration under study by:
$q=\lambda_{2}/\lambda_{1}\equiv b/a$, with $\lambda_{1}>\lambda_{2}$.
The corresponding eigenvectors provide the direction of the principal axes.

\subsection{Projection effects}
Random projections of field galaxies could appear as groups of
galaxies, which however would bare no relation to dynamical entities.
We have performed Monte-Carlo simulations ($N_{sim}=10000$) and found that
sets of 4 points uniformly distributed within a sphere and projected on a 
plane give a nearly Gaussian $q$-distribution with
$\langle q \rangle \simeq 0.6$ and $\sigma_{q} \simeq 0.18$, 
while for sets of 10 points the corresponding values
are $\langle q \rangle \simeq 0.72$ and $\sigma_{q}\simeq 0.13$ respectively, 
significantly larger than the observed case (see Table 1). As an
example we compare in Fig. 4 the axial ratio distribution of our 
groups (having 4 galaxy members) with the corresponding Monte-Carlo random
realizations. The difference is evident.
However, this does not preclude
the possibility of projection effects altering the apparent shape and the
dynamical parameters of real groups. 
\begin{figure}
\mbox{\epsfxsize=10cm \epsffile{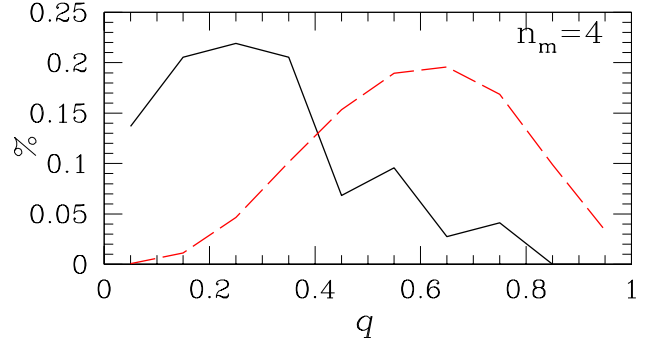}}
\caption{Comparison of the axial ratio distribution of the
 groups with 4 members (continuous line)
with the corresponding distribution of random Monte-Carlo groups
 (dashed line).}
\end{figure}

Ramella, Pisani \& Geller (1997) and Diaferio et
al. (1999) based on either geometrical or N-body simulations and
reproducing the group identification procedure have found that between
70\% and 80\% of the UZC-SSRS2 groups should be real dynamical
entities, while the rest
are expected to be the result of superpositions of field galaxies. 
Such a fraction of false groups, however, could be a
serious problem for studies of the group dynamical
properties, and thus should be investigated in detail.

Indeed, during our eye-balling process of verifying the group shape
measurements, we encountered some cases with ``strange'' positional and 
velocity configurations, which appeared indeed to be the results of projection
effects. This prompt us to check the most probable candidates of being
fake groups, those with a relatively high-velocity dispersion.
One such example is the group UZC269, containing four galaxies,
three of which (forming a straight line)
have $\langle v \rangle \simeq 4450$ km s$^{-1}$ and velocity
dispersion $v_{\sigma}\simeq 200$ km s$^{-1}$, while the fourth
galaxy has a velocity $4 \sigma_v$ lower than $\langle v \rangle$.
Another example is UZC156 with 6 members and an overall
$\sigma_v=647$. However, the total projected size of the group is 
$\approx 2.0$ Mpc while it can be separated in both velocity space and
on the plane of the sky, in two triplets having
$\langle v \rangle= 5200\pm 290$ km s$^{-1}$ and $\langle v \rangle=
4200\pm 195$ km s$^{-1}$.
Other fake cases appeared to contain two independent groups or pairs
of galaxies, with large velocity and/or spatial separations. 

Of course the probability of a group being false is inversely
proportional to the group galaxy membership, $n_m$. Random
projections will affect significantly more the apparent
characteristics (dynamical and morphological) of small
groups rather than large ones 
and for this reason we have decided to exclude from our study
groups with $n_m=3$, exactly due to their {\em
a priori} high probability of being chance projections and 
not real physical systems (see also Focardi \& Kelm 2002). 

Furthermore, we have devised an objective algorithm in an attempt to
single out candidate false groups having $n_m=4$ and 5 (which are those
affected the most). We will present this algorithm in an accompanying
paper, which deals with the dynamics of groups (Tovmassian \& Plionis
2003 {\em in preparation}). Here we only present the
basic assumption on which the algorithm is based, which is
that
the probability of a group being false increases with increasing deviation
of its member velocities from a Gaussian distribution having mean and
variance the observed group values. 
We find in total 20 and 15 groups with $n_m=4$ or 5, respectively that
are probably the result of superpositions of galaxies. Note that these
groups constitute a relatively small fraction ($\sim 14\%$) of the 
sample of 245 groups, less than the expected number of false
groups, according to Ramella et al. (2002). 

We have also tested
our results (presented in section 3)
by including these groups and found that indeed they do
create problems in the inversion from the projected to the intrinsic
3D axial ratio distribution (they cause the inverted distribution for
the prolate case to have unphysical negative values; see section 3).

Since the probability of a group being affected by projections is
inversely proportional to $n_m$, we will present results also 
separately for groups with $n_m=4$ and $4\le n_m\le 10$.

\subsection{Mean group shape parameters}
The summary of the main structure parameters of the different membership
samples of groups is presented in Table 1. The
first and second columns give the group membership, $n_m$, and the 
number $N$ of such groups, respectively, while the third and fourth
columns show the median $q$ and $a$ (major axis) values, together with their
68\% and 32\% quantile values. 
It is evident that the considered UZC-SSRS2 groups are very elongated,
significantly more than what expected from random projections of field
galaxies, giving support to them being real dynamical entities.
Furthermore, among these groups there are many so called chain-like
groups, with $q$ smaller than 0.20-0.30.
The large number of chain-like groups, which also
determines the relatively small median values of $q$, is similar to that of
compact groups, the predominance of which 
was first mentioned by Arp (1973). 


\begin{table}
\caption[]{The median axial ratio, $\bar{q}$,
and the median value of the maximum intergalaxy separation,
$\bar{a}$ (in $h^{-1}$ Mpc), of the different membership
UZGC groups with $cz<5500$ km s$^{-1}$ (after exclusion of 
candidate false groups).} 
\tabcolsep 10pt
\begin{tabular}{ccccc} \\ \hline
 $n_m$& N   & $\langle z \rangle$ & $\bar{q}$   & $\bar{a}$ \\ \hline
 4    & 72  & 0.0115 &$0.27^{+0.06}_{-0.08}$ & $0.54^{+0.11}_{-0.07}$ \\
 5-10 & 108 & 0.0113 &$0.36^{+0.09}_{-0.08}$ & $0.81^{+0.17}_{-0.15}$ \\
 all  & 210 & 0.0113 &$0.36^{+0.11}_{-0.08}$ & $0.77^{+0.26}_{-0.16}$ \\
\hline
\end{tabular}
\end{table}

Note that the median $q$ of all poor groups ($4\le n_m\le 10$) is
$0.33^{+0.09}_{-0.08}$. A certain correlation is apparent
with the median $b/a$ and $a$ increasing with $n_m$. 
The increase of the major axis with $n_m$ could have been due to the
increase with redshift of the group linking parameter (see Ramella
et al. 2002), a fact that induces the systematic increase with $z$ 
of the group size (see Fig. 1). 
However, we have verified that the mean redshift (see 
Table 1) of each group subsample is constant 
and thus the above systematic effect is apparently 
not the cause of the observed increase of $a$ with $z$.

The increase of the group sphericity
could be explained as an indication of a higher degree of
virialization, which is expected to be more rapid in systems
containing more galaxies (mass). 
However, the increase of their major axis, if proven to be true,
is somewhat more perplexing.

\subsection{The projected axial ratio distribution} 
We now proceed in describing our approach of fitting the observed
discrete distribution of axial ratios with the so-called kernel
estimators. 
Here we review the basic steps of the Kernel method, following the
notation of Ryden (1996) but for further
extensive reviews see Silverman (1986), Scott (1992),
Vio et al. (1994) and Tremblay \& Merritt (1995).

Given the sample of axial ratios $q_{1},q_{2},....,q_{N}$ for $N$
groups, the kernel estimate of the frequency distribution is defined
as:
\begin{equation}\label{eq:ker1}
\hat{f}(q)=\frac{1}{Nh} \sum_{i}^{N}\ K\left(\frac{q-q_{i}}{h}\right) \; \; ,
\end{equation}
where $K(t)$ is the kernel function, defined so that 
\begin{equation}\label{eq:ker2}
\int_{-\infty}^{+\infty}\ K(t) {\rm d}t=1 \; \; ,
\end{equation}
and $h$ is the ``kernel width" which determines the balance between
smoothing and noise in the estimated distribution.
In general the value of $h$ is chosen so that the expected value of the
integrated mean square error between the true, $f(q)$, and estimated, 
$\hat f(q)$, distributions, 
$\int_{-\infty}^{+\infty} \left[\hat{f}_{K}(x)-f(x) \right]^{2} {\rm d}x$,
is minimised (cf. Vio et al. 1994; Tremblay \& Merritt 1995).
As it has been shown in different studies the choice of a kernel
function, $K(t)$, among quadratic, quartic and Gaussian forms 
(cf. Tremblay \& Merritt 1995), provide fits that differ trivially only in their
asymptotic efficiencies. We have chosen a Gaussian kernel:
\begin{equation}\label{eq:gaus2}
K(t)=\frac{1}{\sqrt{2\pi}} e^{-t^{2}/2} \; \; .
\end{equation}

In order to obtain physically physically 
acceptable results with $\hat{f}(q)=0$
for $q<0$ and $q>1$, we apply reflective boundary conditions 
(cf. Silverman 1986; Ryden 1996), replacing
the Gaussian kernel with:
\begin{equation}\label{eq:Ker3}
K(q,q_{i},h)=K\!\!\left(\frac{q-q_{i}}{h}\right)+
K\!\!\left(\frac{q+q_{i}}{h}\right)+K\!\!\left(\frac{2-q-q_{i}}{h}\right) \;,
\end{equation}
which also ensures the correct normalization, $\int_{0}^{1} \hat{f}(q) 
{\rm d}q =1$. 

\begin{figure}
\mbox{\epsfxsize=10cm \epsffile{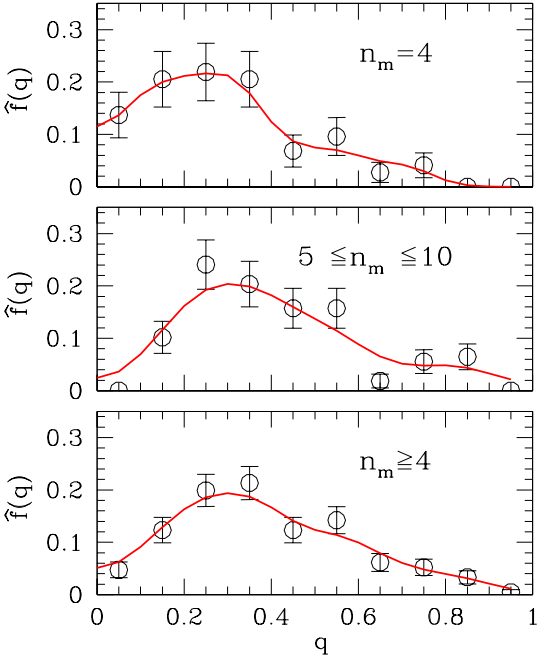}}
\caption{The apparent axial ratio distributions for different group
membership. The solid line is the smooth fit from the nonparametric 
kernel estimator.}
\end{figure}

In Figure 5 we present the projected axial ratio distributions for the
different membership groups (circles), as indicated in the different
panels, with their Poisson 1$\sigma$ error bars, while the 
solid lines shows the
kernel estimate $\hat{f}$ for the appropriate width, $h$.

\section{True Group Shapes}
In order to find the intrinsic axial ratio distribution assuming that
groups are either oblate or prolate spheroids, we use an inversion
method, described below. Although there is no physical justification 
for the restriction to oblate or prolate spheroids,
it greatly simplifies the inversion problem. Furthermore, if groups
are a mixture of
the two spheroidal populations or they have triaxial configurations
then there is no unique inversion (Plionis, Barrow \& Frenk 1991).
\begin{figure}
\mbox{\epsfxsize=10.cm \epsffile{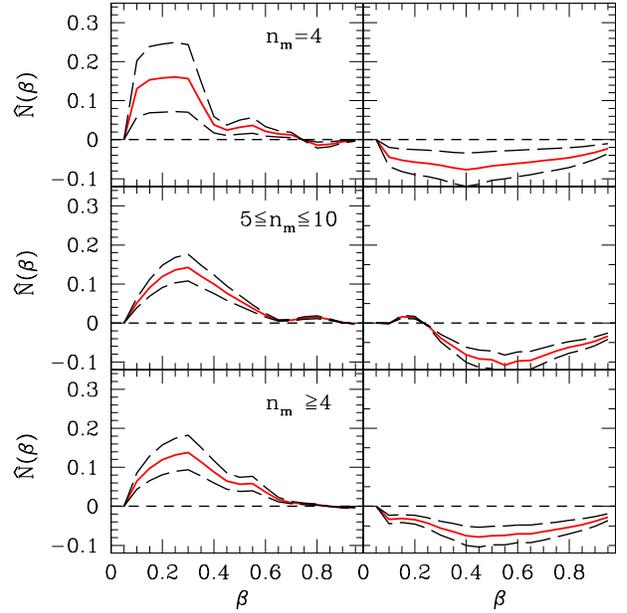}}
\caption{The distribution of the intrinsic poor group axial ratios
(continuous line) and 
its $1 \sigma$ range (broken lines) 
assuming that they are either prolate (left
panel) or oblate (right panel) spheroids.}
\end{figure}

The relation between the apparent and intrinsic axial ratios, is
described by a set of integral equations first investigated by Hubble
(1926).
These are based on the assumptions that the orientations are random
with respect to the line of sight, and that the intrinsic shapes can
be approximated by either oblate or prolate spheroids.
Writing the intrinsic axial ratios as $\beta$ and the estimated distribution
function as $\hat N_o(\beta)$ for oblate spheroids, and $\hat
N_p(\beta)$ for prolate spheroids then the corresponding distribution
of apparent axial ratios is given for the oblate case by:
\begin{equation}\label{eq:apaobl}
\hat{f}(q)=q\int_{0}^{q}\frac{\hat{N}_{\circ}(\beta) {\rm d}\beta}
{(1-q^{2})^{1/2}(q^{2}-\beta^{2})^{1/2}}
\end{equation}
and for the prolate case by:
\begin{equation}\label{eq:apaprol}
\hat{f}(q)=\frac{1}{q^{2}}\int_{0}^{q}\frac{\beta^{2}\hat{N}_{p}(\beta) 
{\rm d}\beta}
{(1-q^{2})^{1/2}(q^{2}-\beta^{2})^{1/2}} \; \; .
\end{equation}
Inverting equations (eq.\ref{eq:apaobl}) and (eq.\ref{eq:apaprol})
gives us the distribution of real axial ratios as a function of the measured
distribution:
\begin{equation}\label{eq:oblate}
\hat{N}_{o}(\beta)=\frac{2\beta (1-\beta^{2})^{1/2}}{\pi} \int_{0}^{\beta}
\ \frac{\rm d}{{\rm d}q}\left(\frac{\hat{f}}{q} \right)\frac{{\rm d}q}
{(\beta^{2}-q^{2})^{1/2}}
\end{equation}
and
\begin{equation}\label{eq:prolate}
\hat{N}_{p}(\beta)
=\frac{2(1-\beta^{2})^{1/2}}{\pi\beta} \int_{0}^{\beta}
\ \frac{\rm d}{{\rm d}q}(q^{2}\hat{f})
\frac{{\rm d}q}{(\beta^{2}-q^{2})^{1/2}} \; \; .
\end{equation}
with $\hat{f}(0)=0$. In order for $\hat{N}_{p}(\beta)$ and $\hat{N}_{o}(\beta)$
to be physically meaningful they should be positive for all
$\beta$'s. 
Following Ryden (1996), we numerically integrate eq.(\ref{eq:oblate})
and eq.(\ref{eq:prolate}) allowing $\hat{N}_{p}(\beta)$ and
$\hat{N}_{o}(\beta)$ to take any value. 
If the inverted distribution of axial ratios has significantly
negative values, a fact which is unphysical, then this can be  
viewed as a strong indication 
that the particular spheroidal model is unacceptable.

In Figure 6 we present the intrinsic group axial
ratio distributions. The oblate model (right panel) is completely
unacceptable since it produces only negative values of the inverted
intrinsic axial ratio distribution.
The UZC-SSRS2 groups shape is represented well only
by that of prolate spheroids which is in agreement with previous studies of
albeit smaller group samples (cf. Oleak et al 1995). 
It is very interesting the fact that
there are almost no groups with true axial ratio, $\beta \magcir 0.6$, ie.,
there are no roundish groups while most of them are extremely
elongated. The most elongated groups are the poorer ones (those with
$n_m=4$) with 85\% having $\beta\mincir 0.4$. We can crudely
approximate the intrinsic prolate axial ratio distribution of all the UZC-SSRS2
groups of our sample (lower left panel of Fig. 6) by a Gaussian having
$\langle \beta \rangle=0.29$ and $\sigma=0.16$ (see Fig. 7).
\begin{figure}
\mbox{\epsfxsize=11cm \epsffile{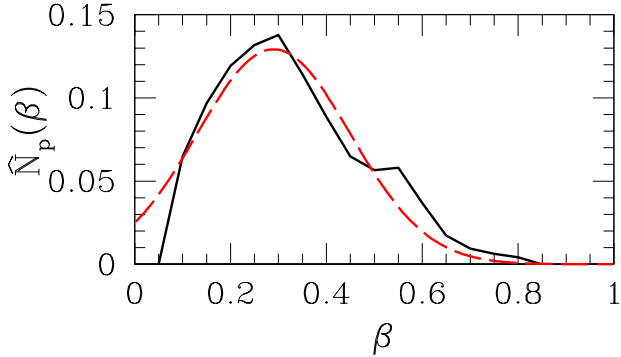}}
\caption{The intrinsic axial ratio distribution (for the prolate case)
of all the UZC-SSRS2 groups of the sample analysed (continuous line)
and the best Gaussian fit (broken line).}
\end{figure}

How do our group shape results compare with those of
clusters of galaxies ? In Figure
8 we present the distribution of intrinsic axial ratio, for the prolate
case (which also is the best model for clusters; see Plionis et al
1991, Basilakos et al 2000), for both the 
UZC-SSRS2 groups and the APM clusters that appear to have no
substructure (which means that they are probably in virial
equilibrium; see Basilakos et al 2000). It is evident that groups are
significantly more elongated than clusters and although there are
quite a few spherical clusters, this is not the case for groups. This
probably implies that 
the cluster distribution is in a more advanced dynamical state,
with violent relaxation having sphericalized the initial flattened
distribution of galaxy members. 
\begin{figure}
\mbox{\epsfxsize=11cm \epsffile{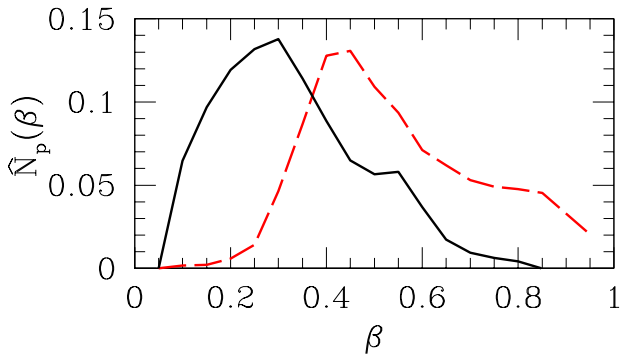}}
\caption{Comparison of the UZC-SSRS2 group (continuous
line) and the APM cluster (broken line; from Basilakos et al 2000) 
intrinsic axial ratio distributions. Note that these distributions are
based on 212 and 405 objects respectively.}
\end{figure}

\section{Discussion \& Conclusions}
We have measured the projected axial ratio of all UZC-SSRS2 
Groups of Galaxies within a volume-limited region ($cz\le 5500$
km/sec) using the moments of the discrete galaxy distribution.

Using the nonparametric kernel procedure we obtain a smooth estimate
of the apparent UZC-SSRS2 group axial-ratio distribution.
The projected axial ratio distribution of the whole group sample 
peaks at $q \simeq 0.33$ with an extended tail towards apparently
spherical groups. Assuming that the UZC-SSRS2 groups constitute a 
homogeneous population of either
oblate or prolate spheroids, we numerically invert the apparent axial ratio
distribution to obtain the corresponding intrinsic one.  The only
acceptable model is provided by that of prolate spheroids having
an intrinsic distribution with $\langle\beta \rangle \simeq 0.29$ and
$\sigma_\beta=0.16$ (if modeled by a Gaussian). This results are in very good 
agreement with the analysis of 95 Shakhbazian compact groups (Oleak et al 1995) and 
shows that generically poor groups of galaxies, compact or not, are
extremely elongated prolate systems, much more than clusters or even
elliptical galaxies.

This result supports the view by which groups form by accretion of galaxies
along larger structures, like filaments, in which they could be
embedded (cf. West 1994).
Such an accretion process would happen preferentially
along the group major axis, which should then be typically aligned
with other nearby structures (groups or clusters). Such an effect is
well documented for clusters (cf. Bingelli 1982; Plionis 1994,
Plionis \& Basilakos 2002 and
references therein), but it has still to be determined for groups of galaxies.

Finally, we would like to comment on the orientation effects that the
extremely elongated and intrinsically prolate nature of groups would create.
When elongated groups are orientated roughly along the line of sight 
they will appear roughly
spherical therefore  having a higher velocity dispersion and smaller sizes 
while elongated groups seen roughly orthogonal to the line of sight
will appear to have smaller velocity dispersions. This would affect
relations like that between the group's
X-ray luminosity - velocity dispersion and thus if not taken
into account, erroneous conclusions could be reached regarding their
dynamical state (see Tovmassian, Yam \& Tiersch 2002;
Plionis \& Tovmassian 2004).

\section* {Acknowledgements}
MP acknowledges funding by the Mexican Government grant
No. CONACyT-2002-C01-39679. SB acknowledges the hospitality of
INAOE where this work was completed.

\end{document}